\documentclass[sigconf]{acmart}

\AtBeginDocument{%
  }

\copyrightyear{2026}
\acmYear{2026}
\setcopyright{cc}
\setcctype{by}
\acmConference[MLCAD '26]{2026 ACM/IEEE International Symposium on Machine Learning for CAD}{September 07--09, 2026}{Jeju Island, Republic of Korea}
\acmBooktitle{2026 ACM/IEEE International Symposium on Machine Learning for CAD (MLCAD '26), September 07--09, 2026, Jeju Island, Republic of Korea}
\acmDOI{10.1145/3831599.3840344}
\acmISBN{979-8-4007-2878-5/2026/09}

\usepackage{amsmath}
\usepackage{xcolor}
\usepackage{graphicx}
\usepackage{multirow}
\usepackage{makecell} 
\usepackage{enumitem}
\usepackage{balance}

\definecolor{propbg}{HTML}{EBF5FB}
\definecolor{critbg}{HTML}{FDEDEC}
\definecolor{evalbg}{HTML}{F0F0F0}
\definecolor{propbg}{HTML}{EBF5FB}
\definecolor{propborder}{HTML}{2980B9}
\definecolor{critbg}{HTML}{FDEDEC}
\definecolor{critborder}{HTML}{E74C3C}
\definecolor{evalbg}{HTML}{F0F0F0}
\definecolor{taggreen}{HTML}{27AE60}
\definecolor{tagred}{HTML}{C0392B}
\definecolor{impgreen}{HTML}{1E8449}

\graphicspath{{figures/}}
\makeatletter
\patchcmd{\@subsection}{\parindent}{\z@}{}{}
\makeatother
\newcolumntype{L}[1]{>{\raggedright\arraybackslash}p{#1}}
\newcommand{\sysname}{TraceDSE}

\definecolor{propbg}{HTML}{E1F5EE}
\definecolor{propfg}{HTML}{04342C}
\definecolor{critbg}{HTML}{FAECE7}
\definecolor{critfg}{HTML}{4A1B0C}
\definecolor{evalbg}{HTML}{F5F5F0}
\definecolor{evalfg}{HTML}{3D3D3A}
\definecolor{tagbg}{HTML}{FAEEDA}
\definecolor{tagfg}{HTML}{854F0B}
\definecolor{impbg}{HTML}{E1F5EE}
\definecolor{impfg}{HTML}{0F6E56}

\renewcommand{\arraystretch}{0.93}
\microtypesetup{expansion=true,stretch=30,shrink=40,step=2}

\begin{document}

%% ─── Title ──────────────────────────────────────────────────
\title[Agentic DSE for AI Inference on Edge SoCs]{Agentic Design Space Exploration for Joint Hardware Configuration Selection and Mapping of AI Inference Workloads on Heterogeneous Edge SoCs}

%% ─── Authors ────────────────────────────────────────────────
%% NOTE: Fill in real author names, ORCIDs, affiliations, and
%% emails. Duplicate the \author block for each author.
\author{Geetha Prasuna Yarramneni$^\S$, Surya Selvam$^\S$, Wilfried Haensch$^\#$$^*$ and Anand Raghunathan$^\S$}
\affiliation{%
  \institution{$^\S$Elmore Family School of ECE, Purdue University, West Lafayette, IN, USA\\ $^\#$Pritzker School of Molecular Engineering, University of Chicago, Chicago, IL, USA\\
  $^*$Argonne National Laboratory, Argonne, IL, USA}
  \country{USA}
}
\begin{comment}
\author{Surya Selvam}
\affiliation{%
  \institution{Purdue University}
  \city{West Lafayette}
  \state{IN}
  \country{USA}}

\author{Wilfried Haensch}
\affiliation{%
  \institution{University of Chicago}
  \city{Chicago}
  \state{IL}
  \country{USA}}

\author{Anand Raghunathan}
\affiliation{%
  \institution{Purdue University}
  \city{West Lafayette}
  \state{IN}
  \country{USA}}
 
\end{comment}

%% Short author list for running headers.
\renewcommand{\shortauthors}{Yarramneni et al.}
\begin{abstract}
Modern edge Systems-on-Chip (SoCs) integrate heterogeneous processing units such as CPUs, GPUs, and NPUs, each with distinct performance and energy characteristics. Deploying AI inference workloads on such platforms under real-time latency and energy constraints requires jointly mapping workloads to processing units (PUs) and configuring each PU (e.g., by selecting the number of active cores and operating frequency). This joint space grows combinatorially, making exhaustive search for the optimal deployment strategy computationally infeasible. Most prior work on design space exploration (DSE) has applied black-box optimization (BBO) methods such as evolutionary search to navigate such vast design spaces. An evaluation (simulation or execution on real hardware) is used to obtain metrics, such as latency and energy, to guide the search. Recently, Large Language Model (LLM)-guided approaches for DSE have gained interest but continue to use sparse feedback similar to BBO approaches. We observe that sparse feedback limits the efficiency of LLM-driven DSE since (i) it does not foster insight into the nature of the design space or the specific reasons that lead a given design choice performing the way it does, and (ii) it does not leverage the full reasoning abilities of LLMs, which are capable of taking richer information from the evaluation of each design point to make higher-quality decisions.\looseness=-1\relax{}

Based on these insights, we present \sysname{}, an agentic DSE flow that performs joint workload mapping and PU configuration selection for AI inference workloads on heterogeneous SoCs. \sysname{} consists of an iterative proposer--critic loop driven by rich feedback in the form of  system execution traces. The LLM proposer agent generates candidate mappings and PU configurations for hardware evaluation. The LLM critic agent, equipped with programmatic trace-analysis tools, analyzes execution traces to identify bottlenecks and suggest targeted refinements. This feedback loop enables deeper insights into each design point, higher-quality decisions, and eventually a more effective search of the design space.\looseness=-1\relax{}
%These recommendations give the proposer visibility into each candidate's execution, enabling it to generate better-performing candidates in the subsequent iterations.

Across four AI inference workloads (models of varying complexity and an application pipeline composed of multiple models) deployed on an Intel Meteor Lake SoC, \sysname{} consistently outperforms two state-of-the-art BBO tools. Specifically, \sysname{} improves Pareto frontier hypervolume by up to 35\% over NSGA-II evolutionary search and by up to 68\% over Bayesian optimization, while requiring ${\sim}$6--9$\times$ fewer hardware evaluations.\looseness=-1\relax{}
\end{abstract}

%% ─── CCS / Keywords ────────────────────────────────────────


\maketitle

\vspace{-4pt}
\section{Introduction}
\label{sec:intro}
\begin{figure}[t]
  \centering
  \includegraphics[width=0.9\columnwidth]{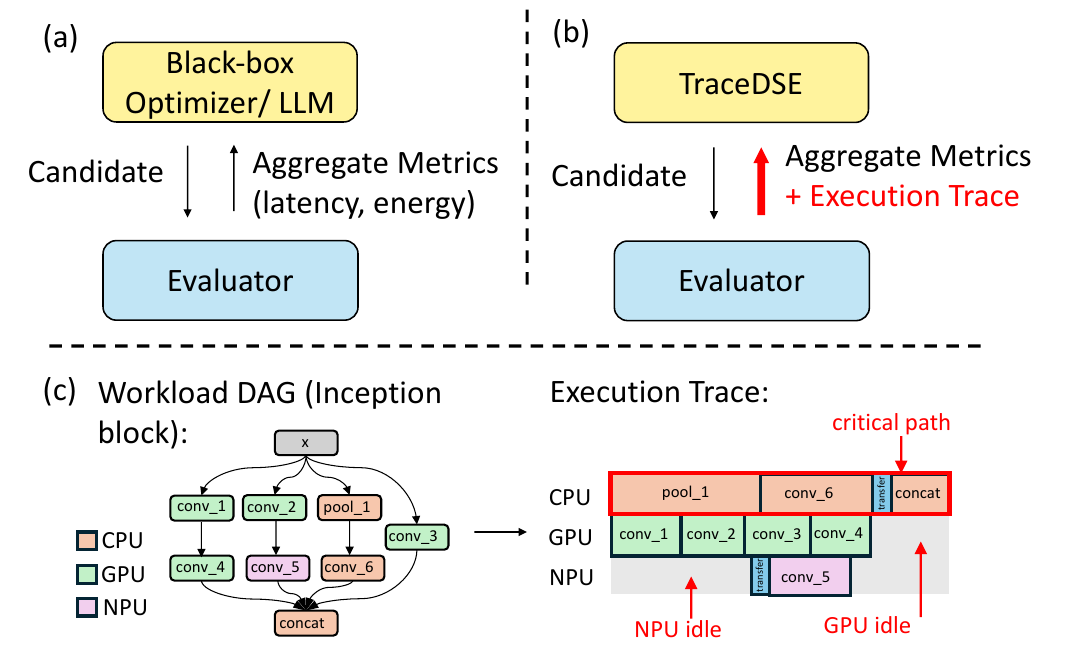}
\caption{(a) Conventional DSE: the optimizer receives only aggregate metrics. (b) TraceDSE: the evaluator also returns an execution trace, enabling targeted refinement. (c) Execution trace of a block of the Inception DNN revealing opportunities that aggregate metrics miss.}
  \Description{Three-part schematic. (a) A conventional design-space-exploration loop in which the optimizer receives only aggregate latency and energy from the evaluator. (b) The TraceDSE loop, in which the evaluator additionally returns an execution trace. (c) A Gantt-style execution trace of one Inception block on CPU, GPU and NPU lanes, showing critical-path nodes on the CPU, two convolutions queued on the GPU, and an idle NPU.}
  \label{fig:motivation}
\end{figure}

AI models are being widely deployed on edge hardware such as mobile phones, wearables, smart glasses, and AR/VR headsets, powering on-device applications across vision (face recognition, gesture and gaze tracking, and computational photography), speech (transcription and translation), and natural language processing (personal assistants)~\cite{chen2019deep}.
These devices typically contain heterogeneous SoCs that integrate multiple PUs---CPUs, GPUs, and NPUs.
Efficiently deploying models onto such SoCs requires jointly navigating two design spaces: the \emph{workload mapping space} and the \emph{PU configuration space}.
\textbf{(1)~Workload mapping space:} A neural network workload is commonly represented as a computational graph, where each node corresponds to a computational kernel, such as a convolution, fully-connected layer or attention computation. Mapping such a  graph onto a heterogeneous SoC requires selecting which PU each node executes on. Each PU could offer a different latency--energy trade-off for the same node. For example, a node that upsamples a $48\times48$ feature map to $192\times192$ and applies a $1\times1$ convolution ($192\to1000$ channels, ${\sim}7$\,GFLOPs) when executed on an Intel Meteor Lake SoC takes 46.7\,ms and 1.40\,J on the GPU, 51.3\,ms and 1.04\,J on the NPU, and 51.5\,ms and 3.81\,J on the CPU---the GPU is 10\% faster but costs 26\% more energy than the NPU, while the CPU matches the NPU's latency at 3.7$\times$ the energy cost. Beyond these node-level tradeoffs, the overall goal is to utilize the collection of PUs to achieve the desired overall latency and energy goals for the full application.
%For example, a convolution node on an Intel Meteor Lake SoC incurs different latency and energy costs on the CPU, the integrated GPU, and the NPU.
\textbf{(2)~PU configuration space:} Each PU exposes runtime-configurable options that further shape performance--energy behavior. For example, on an Intel Meteor Lake SoC, the CPU offers a choice between Performance cores (P-cores) and Efficiency cores (E-cores) with a configurable active core count. The GPU and NPU expose operating frequency control, and the NPU additionally exposes a tile count (number of active compute tiles).\looseness=-1\relax{} 
The cross-product of choices in these two spaces yields a vast and complex design space.
For example, deploying HRNet~\cite{hrnet}, a pose-estimation model with 98 nodes, across 3 PUs on an Intel Meteor Lake SoC---each PU offering 2 configurations---yields a design space with $3^{98} \times 2^{3} \approx 4.6 \times 10^{47}$ candidates. Each candidate requires 30--60 seconds per hardware evaluation, making exhaustive search infeasible.\looseness=-1\relax{}
 
Navigating this design space to identify candidates that achieve the best latency--energy trade-offs requires an efficient search strategy.
Prior iterative DSE methods follow a common paradigm (Figure~\ref{fig:motivation}(a)): an optimizer proposes a candidate (or a population of candidates), an evaluator returns aggregate metrics (e.g., end-to-end latency and energy), and the optimizer uses this feedback to propose the next candidate(s). Black-box optimizers (BBOs) such as genetic algorithms and Bayesian optimization operate within this loop, relying solely on aggregate metrics to guide the search~\cite{spotlight,archgym}. These methods lack the domain knowledge to avoid inferior regions of the design space, leading to slow convergence.
Prior work has addressed this by embedding hand-crafted domain-specific heuristics into these optimizers~\cite{farsi,moham,spotlight}, but such heuristics require manual engineering effort for each new platform or workload.\looseness=-1\relax{}
More recently, LLM-guided approaches have automated this step, leveraging knowledge embedded in pre-trained LLMs to guide the search~\cite{gem5copilot,lemoe,llambo,llmbox,llmdse,idse}. All these methods still rely on aggregate metrics, limiting search efficiency.
%We empirically demonstrate that LLM agents, when limited to aggregate feedback, fail to identify the design choices responsible for a candidate’s underperformance: removing execution-trace feedback degrades Pareto front quality significantly, with up to 73\% higher energy at iso-latency (Section~\ref{subsec:ablation}).

Evaluating a candidate on hardware has the potential to yield far more insight than  aggregate metrics like latency and energy. An execution trace can provide information about the execution time and energy of each node, cross-PU data transfer times, PU utilization, and more. Figure~\ref{fig:motivation}(c) illustrates this on a block from the Inception DNN. The execution trace reveals that the critical-path nodes are mapped to the CPU---making it the bottleneck---and that  nodes \texttt{conv\_2} and \texttt{conv\_3} are queued on the GPU while the NPU remains idle. These insights directly suggest two design choices: remapping queued nodes onto idle PUs to shorten the critical path and increasing the active CPU core count to accelerate the critical-path nodes. These decisions cannot be easily derived from aggregate metrics alone. 

%Execution traces are widely used for performance analysis in heterogeneous systems~\cite{cocoml}, but existing LLM-based DSE methods have not effectively incorporated them. \sysname{} uses and reasons on insights from execution traces -- nodes on the critical path, underutilized PUs, and cross-PU data transfers that stall critical nodes -- to inform DSE decisions and improve the efficiency of LLM-guided DSE.

Deriving actionable insights from a raw execution trace requires analyzing per-node execution behavior and reasoning about which mapping and configuration changes would improve performance. LLMs, with their ability to interpret structured representations---execution traces for performance analysis, or workload graphs for mapping 
decisions---are well-suited to automate this within the DSE loop. These two reasoning tasks are distinct: execution analysis demands detailed per-node traces that grow across iterations, while candidate generation requires a macro view of the design space. This motivates decomposing the DSE loop into two specialized agents: one that analyzes execution traces and recommends improvements, and one that generates new candidates.\looseness=-1\relax{}

The key contributions of this work are as follows:
\begin{itemize}[leftmargin=*, nosep]
  \item We present TraceDSE, an agentic DSE flow for joint mapping and PU configuration selection. 
  \item We propose an iterative proposer--critic loop in which the proposer generates candidates and the critic analyzes their execution traces and recommends targeted refinements to the proposer. To balance refinement with exploration, a search state maintains the Pareto front from the design points explored thus far and exposes under-evaluated regions of the design space to the proposer.
  \item We equip the critic with programmatic trace-analysis tools that provide insight into how individual mapping and configuration choices affect latency and energy, enabling the proposer to act on richer, structured feedback rather than on aggregate metrics alone.
  \item We evaluate TraceDSE on four AI inference workloads (models of varying complexity and an application pipeline composed of multiple models) executed on an Intel Meteor Lake SoC. TraceDSE consistently
outperforms two state-of-the-art BBO tools. Specifically, TraceDSE
improves Pareto frontier hypervolume by up to 35\% over NSGA-II
evolutionary search and by up to 68\% over Bayesian optimization, while requiring ${\sim}$6--9$\times$ fewer hardware evaluations.

\end{itemize}
%% ═════════════════════════════════════════════════════════════
%% 2. RELATED WORK  (~0.5 pages)
%% ═════════════════════════════════════════════════════════════
\begin{figure*}[!t]
  \centering
  \includegraphics[
    width=0.85\textwidth,
    height=0.45\textheight,
    keepaspectratio
  ]{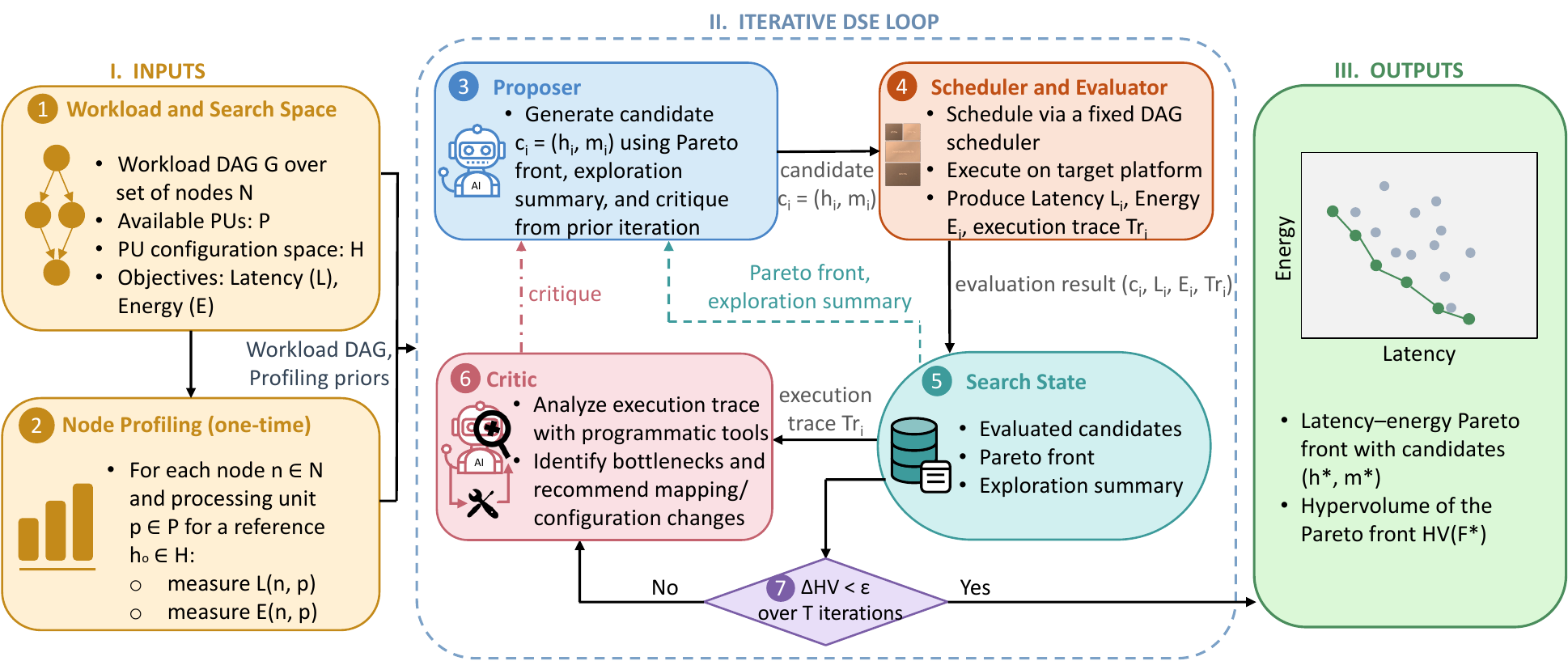}
      \caption{TraceDSE framework overview. The proposer generates candidates using the workload DAG, profiling priors, and prior-iteration feedback. The critic analyzes execution traces with programmatic tools to recommend changes.}
  \Description{Block diagram of proposer--evaluator--critic loop with Pareto tracking and convergence detection.}
  \label{fig:framework}
\end{figure*}

\vspace{-4pt}
\section{Related Work}
\label{sec:related_work}

Existing DSE methods employ different search strategies within an optimizer--evaluator loop.
ArchGym~\cite{archgym} provides a unified framework for comparing search algorithms, including evolutionary algorithms, Bayesian optimization, reinforcement learning, and ant colony optimization, across architectural design spaces.\looseness=-1\relax{}
MOHaM~\cite{moham} uses a multi-objective evolutionary algorithm to jointly search chiplet configurations, chiplet placement, and layer-to-chiplet assignments for multi-DNN deployment.
FARSI~\cite{farsi} augments simulated annealing with architectural heuristics that leverage system-level estimates of compute and memory bottlenecks to guide exploration within the SoC design space.
Spotlight~\cite{spotlight} uses Bayesian optimization for accelerator HW/SW co-design by replacing raw design parameters with derived features, such as PE utilization and estimated DRAM traffic, in the surrogate model.\looseness=-1\relax{}
%In all cases, the optimizer receives only aggregate metrics from the evaluator.

Recent work integrates LLMs into the iterative DSE loop.
gem5 Co-Pilot~\cite{gem5copilot} uses an LLM to retrieve and reason over prior simulation results to guide L2 cache design.
LEMOE~\cite{lemoe} searches RISC-V microarchitecture parameters using multi-objective Bayesian optimization, where an LLM warm-starts the search and enhances the surrogate model through in-context learning.
iDSE~\cite{idse} uses LLM-driven convergent and divergent reasoning to refine HLS directive configurations towards the Pareto front.
LLM-DSE~\cite{llmdse} coordinates multiple LLM agents to iteratively propose and refine HLS directives.
LLAMBO~\cite{llambo} applies LLMs to Bayesian optimization for ML hyperparameter tuning by expressing its core components---warm-starting, candidate sampling, and surrogate modeling---as natural-language prompts.
LLM-Box~\cite{llmbox} uses an LLM to warm-start the search and filter infeasible candidates before handing off to standard multi-objective BBO for mapping LLM inference workloads onto AI accelerator pods.
These methods leverage the prior knowledge embedded in pretrained LLMs to improve search efficiency compared with conventional optimizers; however, the feedback they consume is limited to aggregate metrics. We argue that such an approach does not lend to attributing outcomes to individual design decisions, nor does it provide a basis for reasoning about how to improve the current solution.\looseness=-1\relax{}

\vspace{-4pt}
\section{\sysname{} Methodology}
\label{sec:method}

In this section, we formulate the joint mapping and configuration 
problem (Section~\ref{sec:formulation}), present the \sysname{} 
framework (Section~\ref{sec:framework}), and describe each 
component in detail (Sections~\ref{sec:profiling}-\ref{sec:critic}).

\vspace{-7pt}
\subsection{Problem Formulation}
\label{sec:formulation}
 
We represent the workload as a directed acyclic graph (DAG) $G$, with a node set $N$ and edges that encode data dependencies. The target platform provides a set of heterogeneous PUs, $P = \{p_1, p_2, \ldots, p_{|P|}\}$, with full PU configuration 
space $H = H_1 \times \cdots \times H_{|P|}$, where $H_i$ 
denotes PU-specific parameters for PU $p_i$ (e.g., core type and active core 
count for CPUs, tile allocation for NPUs). A mapping $m: N \rightarrow P$ assigns each node to a PU. Given a mapping $m$ and PU configuration $h \in H$, a  scheduler produces a schedule $S(G, h, m)$ that respects the data dependencies in $G$. We optimize two objectives: the end-to-end execution time (latency) $L(h, m) = \text{Makespan}(S(G, h, m))$, and the energy $E(h, m) = \text{Pow}(h, m) \cdot L(h, m)$, where $\text{Pow}(h, m)$ is the average system power during inference.
The quality of a Pareto front $\mathcal{F}$ is measured by the hypervolume $\text{HV}(\mathcal{F})$~\cite{hypervolume}, which denotes the area in the objective space dominated by $\mathcal{F}$. 
The goal is to discover nondominated configurations that maximize $\text{HV}(\mathcal{F})$ in the latency--energy space.\looseness=-1\relax{} %The resulting search space has size $|P|^{|N|} \times |H|$, where $|P|^{|N|}$ counts all possible node-to-PU mappings and $|H|$ counts all PU configuration choices.

\vspace{-7pt}
\subsection{Framework Overview}
\label{sec:framework}
 
TraceDSE is a closed-loop DSE framework consisting of two LLM agents---a proposer and a critic---connected through an iterative evaluate-analyze-refine loop 
(Figure~\ref{fig:framework}). Before the loop begins, one-time node profiling measures per-node latency and energy on each PU. In each iteration, the proposer generates candidate design points, which are scheduled and evaluated on the target platform, producing aggregate metrics and an execution trace. The critic analyzes this trace using programmatic tools (Table~\ref{tab:critic_tools}) and recommends targeted mapping and configuration changes to the proposer. A search state records evaluated candidates and maintains the Pareto front and exploration summary, which the proposer uses alongside the critic's feedback to generate candidates in the next iteration. The loop terminates when the relative improvement in Pareto-front hypervolume remains below a specified threshold (0.5\% in our experiments) over a specified number of consecutive iterations  (five in our experiments).\looseness=-1\relax{}

\vspace{-7pt}
\subsection{Node Profiling}
\label{sec:profiling}

Before the DSE loop begins, node profiling produces per-node latency and energy measurements on each PU, which we refer to as \emph{profiling priors}. TraceDSE obtains these by running the workload on the target platform once per PU, mapping all nodes to that PU under maximum resource settings. The profiling priors serve two roles: they warm-start the proposer for generating informed initial candidates, and they enable the critic to estimate the end-to-end latency impact of a proposed remapping by substituting a node's profiled latency on the proposed PU without additional hardware evaluation.\looseness=-1\relax{}

\vspace{-7pt}
\subsection{Proposer Agent}
\label{sec:proposer}
 
The proposer generates a batch of candidate design points in each
iteration. Each candidate consists of a mapping, a PU configuration,
and a reasoning statement explaining its design choices.

To generate these candidates, the proposer must determine a mapping of nodes to PUs and a configuration for each PU. We
provide four inputs that inform these decisions: the
\textit{workload DAG}, \textit{PU configuration space},
\textit{optimization objectives}, and \textit{profiling priors}
(Section~\ref{sec:profiling}). The workload DAG lets the proposer
exploit concurrency by distributing independent nodes across PUs
while respecting ordering constraints. The profiling priors let
the proposer avoid clearly inefficient PU choices when
constructing initial mappings.

In subsequent iterations, the proposer receives three additional
inputs to guide its search. The
\textit{Pareto front} captures the best latency--energy trade-offs
discovered at any point in the search, guiding the proposer toward regions in the
objective space where new candidates would most improve
hypervolume. The \textit{exploration summary} records which mapping
and configuration regions have been evaluated, so the proposer
avoids revisiting them. The \textit{critique} from the prior
iteration provides per-candidate feedback identifying which design
choices drove improvements in the metrics, along with recommended changes to the mapping and PU configuration, which the proposer uses to inform
its next batch of candidates.\looseness=-1\relax{}

\vspace{-7pt}
\subsection{Scheduler \& Evaluator}
\label{sec:evaluator}

Given a candidate design point, the scheduler assigns nodes to PUs according to the candidate's mapping and schedules them using a DAG scheduler, Heterogeneous Earliest Finish Time (HEFT)~\cite{heft}, respecting data dependencies in the workload DAG. The evaluator executes the scheduled workload on the target platform and produces the end-to-end latency, energy, per-PU energy breakdown, and an execution trace---a per-node record of the assigned PU, start and end timestamps, compute time, and inter-PU data transfer delays. All evaluation results are stored in the search state.\looseness=-1\relax{}

\vspace{-7pt}
\subsection{Critic Agent}
\label{sec:critic}

The critic analyzes each candidate's execution trace and produces
a critique consisting of three parts: \textit{per-candidate
feedback} identifying which design choices drove the measured
metrics, so the proposer knows what to preserve and what to
change; \textit{observations} synthesized across current and prior
iterations that capture recurring patterns---for example, that a
particular node-to-PU mapping consistently underperforms; and
\textit{recommended changes} to each candidate's mapping and
configuration, informing the proposer's next batch of candidates.

To distill the execution trace into actionable feedback, the
critic uses six programmatic tools
(Table~\ref{tab:critic_tools}) that it autonomously selects and
invokes---for example, first extracting the critical path to
identify bottleneck nodes, then inspecting those nodes to
determine which alternative PU would reduce their latency. Using
the profiling priors (Section~\ref{sec:profiling}), the critic can
also estimate the latency impact of a proposed remapping,
screening out changes unlikely to improve performance---though
these estimates do not capture runtime contention that arises only
during hardware execution.\looseness=-1\relax{}
\begin{table}[t]
\centering\small
\caption{Critic tool interface. The critic autonomously selects tools to analyze an execution trace.}
\label{tab:critic_tools}
\renewcommand{\arraystretch}{1.12}
\setlength{\tabcolsep}{3pt}
\begin{tabular}{@{}L{0.31\columnwidth}L{0.64\columnwidth}@{}}
\toprule
\textbf{Tool} & \textbf{Returns} \\
\midrule
Critical-path extraction &
Nodes on the critical path with each node's latency and share of the total latency \\[2pt]

Latency bottleneck ranking &
Top-$k$ nodes ranked by descending latency with estimated speedup on alternative PUs \\[2pt]

PU utilization and idle-interval analysis &
Per-PU utilization and idle intervals \\[2pt]

Node inspection &
Detailed breakdown of a single node's execution, including compute time, scheduling and communication delays, and estimated latency of the node on each PU \\[2pt]

Energy breakdown &
Energy consumed by each PU as a percentage of total system energy \\[2pt]

Latency estimation &
Predicted latency after a proposed mapping or configuration change \\
\bottomrule
\end{tabular}
\end{table}

%% ═════════════════════════════════════════════════════════════
%% 4. EXPERIMENTAL SETUP  (~0.5 pages)
%% ═════════════════════════════════════════════════════════════

\begin{figure*}[t]
  \centering
    \includegraphics[
    width=0.85\textwidth,
    height=0.45\textheight,
    keepaspectratio
  ]{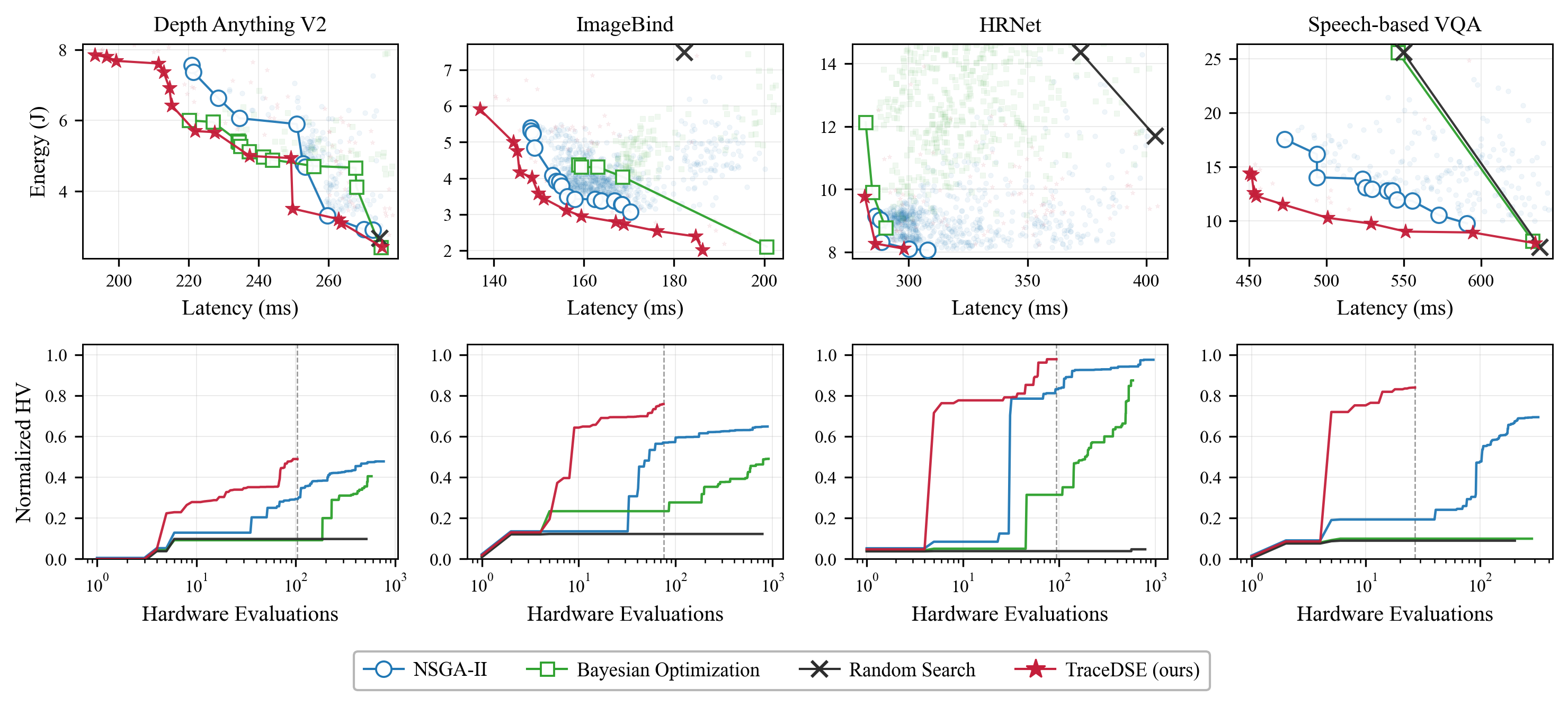}
  \caption{Pareto fronts (top) and normalized hypervolume vs.\ hardware evaluations (bottom, log scale). Faded points are dominated. The dashed line marks TraceDSE's termination; baselines have completed only 11--18\% of their evaluations.}
  \label{fig:pareto_fronts}
  \Description{Visualization of Pareto fronts and normalized hypervolume versus hardware evaluations for four workloads on Intel Meteor Lake.}
\end{figure*}

\vspace{-4pt}
\section{Experimental Setup}
\label{sec:experimental_setup}

We evaluated \sysname{} on an Intel Meteor Lake edge SoC comprising a CPU (6 P-cores, 8 E-cores, 2 low-power E-cores), an integrated GPU, and a 2-tile NPU.
We use OpenVINO~\cite{openvino} as the inference runtime and Gemini~3.1 Pro Preview~\cite{gemini31pro} (temperature 1.0, thinking level \textit{high}) as the LLM for both the proposer and the critic agents.

Each node in the execution trace is mapped to one of three PUs: CPU, GPU, or NPU.
The PU configuration space enumerates valid combinations of five parameters: CPU cores $\in \{1, \ldots, 16\}$, CPU core type $\in$ \{P-core, E-core\}, hyperthreading $\in$ \{on, off\}, NPU turbo $\in$ \{on, off\}, and NPU tile count $\in$ \{1, 2\}.

We evaluate four workloads with varying degrees of parallelism:
Depth Anything V2~\cite{dav2} (28 nodes, mostly sequential),
HRNet~\cite{hrnet} (98 nodes, repeated fork--join parallelism),
ImageBind~\cite{imagebind} (72 nodes, four parallel encoder chains fanning into a joint embedding),
and a speech-based Visual Question Answering (VQA) pipeline combining Whisper-Small~\cite{whisper}, CLIP ViT-L/14~\cite{clip}, and Qwen2.5-3B~\cite{qwen25} (71 nodes, parallel audio and vision encoders fanning into an LLM decoder).\looseness=-1\relax{}

We report the median latency across 50 inference runs after 5 warm-up runs.
Energy per inference is computed as $(P_{\text{avg}} - P_{\text{idle}}) \times \text{latency}$, where $P_{\text{avg}}$ and $P_{\text{idle}}$ are sampled via HWiNFO during inference and when the system is idle, respectively. $P_{\text{avg}}$ is polled from HWiNFO's shared-memory sensors every 0.3\,s (250\,ms hardware refresh) across each configuration's active window of 50 repeated inferences (8--30\,s, 27--94 samples), so sub-second per-inference latency is amortized over tens of samples rather than read from a single sample. Across re-measurements of all the evaluated candidates, the coefficient of variation is 0.3--10.8\% for energy (median 2.8\%) and 0.16--1.2\% for latency.\looseness=-1\relax{}

We compare \sysname{} against three baselines:
random search,
Non-dominated Sorting Genetic Algorithm II (NSGA-II)~\cite{nsga2} (population size 16, uniform crossover),
and multi-objective Bayesian optimization using Tree-structured Parzen Estimators (TPE)~\cite{tpe}, which natively supports the mixed continuous-categorical parameter space.
Each baseline is run for a fixed 12-hour wall-clock budget on the target SoC.
All methods start from the same three seed candidates (all nodes on a single PU, maximum resource settings), counted toward each method's budget.
We normalize objectives per benchmark---using the maximum latency and maximum energy across all methods' Pareto-optimal candidates as the reference point---and report $\widetilde{HV} \in [0, 1]$.\looseness=-1\relax{} 
\vspace{0pt}
\section{Results}
\label{sec:results}
We next present results of our experiments evaluating TraceDSE.

\vspace{-7pt}
\subsection{Pareto Front Quality and Evaluation Efficiency}

\sysname{} achieves the highest hypervolume on all four benchmarks while requiring 5.6--9.0$\times$ fewer evaluations (Table~\ref{tab:tracedse_results}). At iso-latency, \sysname{} achieves 20--32\% lower energy than the strongest baseline on ImageBind and Speech-based VQA (Figure~\ref{fig:pareto_fronts}, top). At iso-energy, TraceDSE achieves 7--18\% lower latency on ImageBind, HRNet, and Speech-based VQA. On three of four benchmarks, TraceDSE discovers low-latency Pareto-front extremes unreached by any baseline---extending 7--14\% beyond the fastest baseline---by exploiting the critic's visibility into idle PU intervals and critical-path bottlenecks.

TraceDSE's advantage is larger when the workload offers more opportunities for concurrent execution across PUs. On HRNet, which has repetitive fork--join parallelism, TraceDSE is only marginally better than NSGA-II (1.1\% hypervolume gap). On ImageBind and Speech-based VQA, where independent encoder chains can be distributed across PUs, the gap widens to 25--29\%. The convergence plots (Figure~\ref{fig:pareto_fronts}, bottom) further illustrate the sample efficiency of trace-driven feedback: TraceDSE reaches near-final hypervolume within 10--30 evaluations and terminates with a higher hypervolume than any baseline achieves after its full 12-hour budget. Of TraceDSE's total wall-clock time, hardware evaluation accounts for 60--65\%, critic LLM inference for 25--35\%, and proposer LLM inference for 4--13\%; evaluation on the target SoC remains the dominant cost.\looseness=-1\relax{}

\begin{table}[t]
\centering\small
\caption{Pareto hypervolume and search time on Intel Meteor Lake. $N_\text{eval}$: hardware evaluations; $\widetilde{HV}$: normalized hypervolume; $t$: wall-clock time. Parentheses: improvement over the strongest baseline.}
\label{tab:tracedse_results}
\setlength{\tabcolsep}{3pt}
\begin{tabular}{llrrr}
\toprule
Benchmark & Method & $N_{\text{eval}}$\,($\downarrow$) & $\widetilde{HV}$\,($\uparrow$) & $t$\,(hrs)\,($\downarrow$) \\
\midrule
\multirow{4}{*}{\parbox{1.6cm}{\raggedright Depth Anything V2}}
 & NSGA-II & 580 & 0.362 & 12.0 \\
 & Bayesian Opt. & 579 & 0.395 & 12.0 \\
 & Random Search & 388 & 0.063 & 12.0 \\
 & \textbf{TraceDSE} & \textbf{104}\,{\footnotesize(5.6$\times$)} & \textbf{0.489}\,{\footnotesize(+24\%)} & \textbf{3.3}\,{\footnotesize(3.6$\times$)} \\
\midrule
\multirow{4}{*}{\parbox{1.6cm}{ImageBind}}
 & NSGA-II & 671 & 0.606 & 12.0 \\
 & Bayesian Opt. & 687 & 0.450 & 12.0 \\
 & Random Search & 593 & 0.018 & 12.0 \\
 & \textbf{TraceDSE} & \textbf{75}\,{\footnotesize(8.9$\times$)} & \textbf{0.758}\,{\footnotesize(+25\%)} & \textbf{2.1}\,{\footnotesize(5.7$\times$)} \\
\midrule
\multirow{4}{*}{\parbox{1.6cm}{HRNet}}
 & NSGA-II & 723 & 0.966 & 12.0 \\
 & Bayesian Opt. & 587 & 0.879 & 12.0 \\
 & Random Search & 594 & 0.033 & 12.0 \\
 & \textbf{TraceDSE} & \textbf{94}\,{\footnotesize(7.7$\times$)} & \textbf{0.977}\,{\footnotesize(+1\%)} & \textbf{2.5}\,{\footnotesize(4.8$\times$)} \\
\midrule
\multirow{4}{*}{\parbox{1.6cm}{\raggedright Speech-based VQA}}
 & NSGA-II & 242 & 0.651 & 12.0 \\
 & Bayesian Opt. & 215 & 0.104 & 12.0 \\
 & Random Search & 152 & 0.096 & 12.0 \\
 & \textbf{TraceDSE} & \textbf{27}\,{\footnotesize(9.0$\times$)} & \textbf{0.839}\,{\footnotesize(+29\%)} & \textbf{2.3}\,{\footnotesize(5.2$\times$)} \\
\bottomrule
\end{tabular}
\end{table}

\begin{table}[t]
\centering\small
\caption{Ablation study across four benchmarks. All variants use
Gemini 3.1 Pro Preview. 
$\Delta$ is the difference in $\widetilde{HV}$ from the full system.}
\label{tab:ablation}
\setlength{\tabcolsep}{3pt}
\begin{tabular}{llrrr}
\toprule
Benchmark & Variant & $\widetilde{HV}$ & $\Delta$ \\
\midrule
\multirow{4}{*}{\parbox{1.6cm}{\raggedright Depth Anything V2}}
 & Full system  & 0.489 & --- \\
 & No trace feedback & 0.414 & $-$0.074 \\
 & No critic  & 0.467 & $-$0.022 \\
 & No search state  & 0.448 & $-$0.041 \\
 & No workload DAG & 0.484 & $-$0.005 \\
\midrule
\multirow{4}{*}{\parbox{1.6cm}{ImageBind}}
 & Full system & 0.758 & --- \\
 & No trace feedback  & 0.647 & $-$0.111 \\
 & No critic  & 0.749 & $-$0.009 \\
 & No search state  & 0.737 & $-$0.021 \\
 & No workload DAG & 0.647 & $-$0.111 \\
\midrule
\multirow{4}{*}{\parbox{1.6cm}{HRNet}}
 & Full system  & 0.977 & --- \\
 & No trace feedback  & 0.901 & $-$0.076 \\
 & No critic  & 0.863 & $-$0.114 \\
 & No search state  & 0.937 & $-$0.040 \\
 & No workload DAG & 0.976 & $-$0.001 \\
\midrule
\multirow{4}{*}{\parbox{1.6cm}{\raggedright Speech-based VQA}}
 & Full system  & 0.839 & --- \\
 & No trace feedback  & 0.773 & $-$0.066 \\
 & No critic  & 0.755 & $-$0.084 \\
 & No search state & 0.837 & $-$0.002 \\
 & No workload DAG & 0.733 & $-$0.106 \\
\bottomrule
\end{tabular}
\end{table}

\vspace{-7pt}
\subsection{Ablation Study}
\label{subsec:ablation}

We ablate four design choices of TraceDSE across all four benchmarks (Table~\ref{tab:ablation}). 

\noindent\textbf{Removing execution trace feedback.}
Restricting the critic to aggregate latency and energy is the only ablation that degrades hypervolume substantially on every benchmark (0.066--0.111) and yields the largest mean drop (0.082); the remaining components are workload-dependent---the critic matters most on HRNet ($-$0.114) and the workload DAG on Speech-based VQA ($-$0.106).\looseness=-1\relax{} At iso-latency, energy increases by up to 73\% on ImageBind. On two of four benchmarks, the full system achieves low-latency Pareto-front extremes that the ablated system cannot reach---12\% lower on ImageBind and 15\% lower on DAv2---as the critic loses visibility into per-node performance bottlenecks and can no longer recommend targeted remappings to reduce critical-path latency.\looseness=-1\relax{}

\noindent\textbf{Removing the critic.}
Passing the raw execution trace directly to the proposer, bypassing the critic and its programmatic tools, degrades hypervolume on all four benchmarks. As raw traces accumulate across iterations, the proposer's input context grows from approximately 10K to over 250K tokens, degrading the quality of the LLM's responses. For Speech-based VQA, the proposer produces increasingly redundant proposals, resulting in lower hypervolume than even the aggregate-only variant.  These results show that the proposer cannot effectively reason over raw trace data and that the critic's structured analysis is necessary to convert raw traces into feedback the proposer can act on.\looseness=-1\relax{}

\noindent\textbf{Removing the search state.}
Removing the Pareto front and exploration summary reduces hypervolume on HRNet and DAv2 but has limited effect on ImageBind and Speech-based VQA. On ImageBind, the ablated system's Pareto front is narrower, failing to reach the low-energy region the full system discovers (12\% higher energy at the energy-optimal extreme). The contrast reflects workload structure: HRNet's repetitive fork-join topology implies the same nodes dominate the critical path across iterations, so without the exploration summary the proposer repeatedly refines the same region; on ImageBind and Speech-based VQA, the bottleneck naturally shifts between independent encoder chains across iterations, exposing different design space regions without explicit tracking.\looseness=-1\relax{}

\noindent\textbf{Removing the workload DAG from the proposer.}
Replacing the workload DAG in the proposer's prompt with unordered node names that do not reveal dependencies among them reduces hypervolume significantly on ImageBind and Speech-based VQA but has no measurable effect on HRNet and DAv2. Unlike the trace ablation, which uniformly degrades the front, removing the DAG causes the system to miss specific Pareto-front extremes while remaining competitive elsewhere. On ImageBind, the ablated system uses 18--49\% more energy at iso-latency and the full system extends 8\% further into the low-energy region. On Speech-based VQA, the full system extends 4\% further into the low-latency region. Reaching these extremes requires the proposer to reason about which nodes can execute concurrently on separate PUs---reasoning that depends on the DAG. DAv2 is mostly sequential, leaving little parallel structure to exploit; HRNet's topology is repetitive across stages, so the critic's trace
analysis compensates.\looseness=-1\relax{}

We present a detailed case study in Appendix~\ref{app:case_study} tracing the proposer–critic loop across 12 iterations on HRNet: the critic identifies a terminal-node bottleneck invisible to aggregate metrics, and the proposer combines this with DAG-level reasoning about concurrent execution to produce a 15.4\% latency reduction.

\vspace{-7pt}
\subsection{Robustness to LLM Choice and Stochasticity}

To evaluate whether TraceDSE's performance is sensitive to LLM sampling nondeterminism or the choice of LLM, we vary both independently.
Across five independent runs of TraceDSE on ImageBind (Gemini 3.1 Pro Preview, temperature~1.0), hypervolume ranges from 0.744 to 0.764 ($\sigma = 0.009$), with 708K\,$\pm$\,88K input and 57K\,$\pm$\,14K output tokens at \$4.46\,$\pm$\,0.65 per run, indicating stable performance across runs. Repeating the five-run protocol on the remaining three benchmarks yields run-to-run variation of the same order.\looseness=-1\relax{}
Table~\ref{tab:llm_comparison} reports TraceDSE results on ImageBind using four different LLMs. All four achieve hypervolume between 0.749 and 0.791---a spread comparable to the run-to-run variance of a single LLM---despite up to 5.3$\times$ differences in cost and 1.9$\times$ in wall-clock time. This indicates that TraceDSE's performance is driven by the proposer--critic architecture and trace-analysis tools rather than by the capability of any individual LLM.\looseness=-1\relax{}
\begin{table}[t]
\centering\small 
\setlength{\tabcolsep}{4pt}
\caption{LLM comparison on ImageBind. All runs use the
full TraceDSE flow, differing only in
the LLM.}
\label{tab:llm_comparison}
\begin{tabular}{lrrrr}
\toprule
LLM & $\widetilde{HV}$ & Tokens (in/out) & $t$ (hrs) & Cost (\$) \\
\midrule
Gemini 3.1 Pro Preview & 0.758 & 737K / 55K & 2.1 & 4.52 \\
Claude Opus 4.7 & \textbf{0.791} & 913K / 138K & 2.6 & 24.00 \\
GPT-5.5 & 0.777 & 422K / 124K & 2.2 & 3.90 \\
Qwen3.6 Max Preview & 0.749 & 214K / 88K & 3.9 & 1.19\\
\bottomrule
\end{tabular}
\end{table}

%%% ===== 6. CONCLUSION ===== %%%

\vspace{-4pt}
\section{Conclusion}
\label{sec:conclusion}

Deploying AI inference workloads on heterogeneous edge SoCs requires jointly optimizing workload mapping and PU configuration across a combinatorially large design space.
However, conventional DSE methods treat the hardware as a black box, relying solely on aggregate performance metrics without exploiting the rich structured information available in execution traces.
To address this, we present 
TraceDSE, an agentic DSE flow that analyzes execution 
traces with programmatic tools to convert each hardware 
evaluation into actionable feedback for guiding the search.
On four single-model and multi-model neural network inference workloads deployed on an Intel Meteor Lake edge SoC, TraceDSE consistently identifies higher-quality candidates and exceeds the Pareto hypervolume of NSGA-II by up to 35\% and Bayesian optimization by up to 68\%, while requiring ${\sim}$6--9$\times$ fewer hardware evaluations.
%Ablation results validate the separation of trace analysis from candidate generation; on larger workloads, exposing raw traces directly to the proposer can degrade performance relative to aggregate-only feedback.
%These results demonstrate that structured execution feedback substantially improves the efficiency of DSE, a principle applicable to other design optimization problems where detailed performance analyses are available.

%% ─── Acknowledgments (omit for anonymous submission) ────────
\begin{acks}
%%This work was supported by \grantsponsor{}{[Funding agency]}{} under Grant No.~\grantnum{}{[XXXX]}.
This material is based upon work supported by the U.S. Department of Energy, Office of Science, Basic Energy Sciences, as part of The Extreme Lithography \& Materials Innovation Center (ELMIC), a Microelectronics Science Research Center (MSRC), under Contract No. DE-AC02-06CH11357. We also acknowledge the support for the Microelectronics Seed Model Team effort from the Transformational AI Models Consortium (ModCon) as part of the DOE Genesis Mission.

%The submitted manuscript has been created by UChicago Argonne, LLC, Operator of Argonne National Laboratory (``Argonne''). Argonne, a U.S. Department of Energy Office of Science laboratory, is operated under Contract No. DE-AC02-06CH11357. The U.S. Government retains for itself, and others acting on its behalf, a paid-up nonexclusive, irrevocable worldwide license in said article to reproduce, prepare derivative works, distribute copies to the public, and perform publicly and display publicly, by or on behalf of the Government. The Department of Energy will provide public access to these results of federally sponsored research in accordance with the DOE Public Access Plan. \url{http://energy.gov/downloads/doe-public-access-plan}

\end{acks}

%% ─── References (do NOT count toward 6-page limit) ──────────
\bibliographystyle{ACM-Reference-Format}
\newpage
\bibliography{references}


\begin{thebibliography}{23}

%%% the Web address.

\ifx \showCODEN    \undefined \def \showCODEN     #1{\unskip}     \fi
\ifx \showISBNx    \undefined \def \showISBNx     #1{\unskip}     \fi
\ifx \showISBNxiii \undefined \def \showISBNxiii  #1{\unskip}     \fi
\ifx \showISSN     \undefined \def \showISSN      #1{\unskip}     \fi
\ifx \showLCCN     \undefined \def \showLCCN      #1{\unskip}     \fi
\ifx \shownote     \undefined \def \shownote      #1{#1}          \fi
\ifx \showarticletitle \undefined \def \showarticletitle #1{#1}   \fi
\ifx \showURL      \undefined \def \showURL       {\relax}        \fi
% invisible to TeX
\providecommand\bibfield[2]{#2}
\providecommand\bibinfo[2]{#2}
\providecommand\natexlab[1]{#1}
\providecommand\showeprint[2][]{arXiv:#2}

\bibitem[Bergstra et~al\mbox{.}(2011)]%
        {tpe}
\bibfield{author}{\bibinfo{person}{James Bergstra}, \bibinfo{person}{R{\'e}mi
  Bardenet}, \bibinfo{person}{Yoshua Bengio}, {and} \bibinfo{person}{Bal{\'a}zs
  K{\'e}gl}.} \bibinfo{year}{2011}\natexlab{}.
\newblock \showarticletitle{Algorithms for Hyper-Parameter Optimization}. In
  \bibinfo{booktitle}{\emph{Advances in Neural Information Processing
  Systems}}, Vol.~\bibinfo{volume}{24}.
\newblock


\bibitem[Boroujerdian et~al\mbox{.}(2023)]%
        {farsi}
\bibfield{author}{\bibinfo{person}{Behzad Boroujerdian}, \bibinfo{person}{Ying
  Jing}, \bibinfo{person}{Devashree Tripathy}, \bibinfo{person}{Amit Kumar},
  \bibinfo{person}{Lavanya Subramanian}, \bibinfo{person}{Luke Yen},
  \bibinfo{person}{Vincent Lee}, \bibinfo{person}{Vivek Venkatesan},
  \bibinfo{person}{Amit Jindal}, \bibinfo{person}{Robert Shearer}, {and}
  \bibinfo{person}{Vijay Janapa~Reddi}.} \bibinfo{year}{2023}\natexlab{}.
\newblock \showarticletitle{{FARSI}: An Early-Stage Design Space Exploration
  Framework to Tame the Domain-Specific System-on-Chip Complexity}.
\newblock \bibinfo{journal}{\emph{ACM Transactions on Embedded Computing
  Systems}} \bibinfo{volume}{22}, \bibinfo{number}{2}, Article
  \bibinfo{articleno}{31} (\bibinfo{year}{2023}), \bibinfo{numpages}{35}~pages.
\newblock
\href{https://doi.org/10.1145/3544016}{doi:\nolinkurl{10.1145/3544016}}


\bibitem[Chen and Ran(2019)]%
        {chen2019deep}
\bibfield{author}{\bibinfo{person}{Jiasi Chen} {and} \bibinfo{person}{Xukan
  Ran}.} \bibinfo{year}{2019}\natexlab{}.
\newblock \showarticletitle{Deep Learning With Edge Computing: A Review}.
\newblock \bibinfo{journal}{\emph{Proc. IEEE}} \bibinfo{volume}{107},
  \bibinfo{number}{8} (\bibinfo{year}{2019}), \bibinfo{pages}{1655--1674}.
\newblock
\href{https://doi.org/10.1109/JPROC.2019.2921977}{doi:\nolinkurl{10.1109/JPROC.2019.2921977}}


\bibitem[Das et~al\mbox{.}(2024)]%
        {moham}
\bibfield{author}{\bibinfo{person}{Abhijit Das}, \bibinfo{person}{Enrico
  Russo}, {and} \bibinfo{person}{Maurizio Palesi}.}
  \bibinfo{year}{2024}\natexlab{}.
\newblock \showarticletitle{Multi-Objective Hardware-Mapping Co-Optimisation
  for Multi-DNN Workloads on Chiplet-Based Accelerators}.
\newblock \bibinfo{journal}{\emph{IEEE Trans. Comput.}} \bibinfo{volume}{73},
  \bibinfo{number}{8} (\bibinfo{year}{2024}), \bibinfo{pages}{1883--1898}.
\newblock
\href{https://doi.org/10.1109/TC.2024.3386067}{doi:\nolinkurl{10.1109/TC.2024.3386067}}


\bibitem[Deb et~al\mbox{.}(2002)]%
        {nsga2}
\bibfield{author}{\bibinfo{person}{Kalyanmoy Deb}, \bibinfo{person}{Amrit
  Pratap}, \bibinfo{person}{Sameer Agarwal}, {and} \bibinfo{person}{T.
  Meyarivan}.} \bibinfo{year}{2002}\natexlab{}.
\newblock \showarticletitle{A Fast and Elitist Multiobjective Genetic
  Algorithm: {NSGA-II}}.
\newblock \bibinfo{journal}{\emph{IEEE Transactions on Evolutionary
  Computation}} \bibinfo{volume}{6}, \bibinfo{number}{2}
  (\bibinfo{year}{2002}), \bibinfo{pages}{182--197}.
\newblock


\bibitem[Fu et~al\mbox{.}(2025)]%
        {gem5copilot}
\bibfield{author}{\bibinfo{person}{Zuoming Fu}, \bibinfo{person}{Alex Manley},
  {and} \bibinfo{person}{Mohammad Alian}.} \bibinfo{year}{2025}\natexlab{}.
\newblock \showarticletitle{{gem5} Co-Pilot: {AI} Assistant Agent for
  Architectural Design Space Exploration}. In \bibinfo{booktitle}{\emph{3rd
  Workshop on Computer Architecture Modeling and Simulation (CAMS), co-located
  with MICRO}}.
\newblock
\newblock
\shownote{arXiv:2510.19577}.


\bibitem[Girdhar et~al\mbox{.}(2023)]%
        {imagebind}
\bibfield{author}{\bibinfo{person}{Rohit Girdhar}, \bibinfo{person}{Alaaeldin
  El-Nouby}, \bibinfo{person}{Zhuang Liu}, \bibinfo{person}{Mannat Singh},
  \bibinfo{person}{Kalyan~Vasudev Alwala}, \bibinfo{person}{Armand Joulin},
  {and} \bibinfo{person}{Ishan Misra}.} \bibinfo{year}{2023}\natexlab{}.
\newblock \showarticletitle{ImageBind: One Embedding Space To Bind Them All}.
  In \bibinfo{booktitle}{\emph{IEEE/CVF Conference on Computer Vision and
  Pattern Recognition (CVPR)}}. \bibinfo{pages}{15180--15190}.
\newblock


\bibitem[{Google DeepMind}(2026)]%
        {gemini31pro}
\bibfield{author}{\bibinfo{person}{{Google DeepMind}}.}
  \bibinfo{year}{2026}\natexlab{}.
\newblock \bibinfo{title}{Gemini 3.1 Pro: A Smarter Model for Your Most Complex
  Tasks}.
\newblock
  \bibinfo{howpublished}{\url{https://blog.google/innovation-and-ai/models-and-research/gemini-models/gemini-3-1-pro/}}.
\newblock


\bibitem[{Intel Corporation}(2024)]%
        {openvino}
\bibfield{author}{\bibinfo{person}{{Intel Corporation}}.}
  \bibinfo{year}{2024}\natexlab{}.
\newblock \bibinfo{title}{{OpenVINO} Toolkit}.
\newblock \bibinfo{howpublished}{\url{https://docs.openvino.ai}}.
\newblock


\bibitem[Krishnan et~al\mbox{.}(2023)]%
        {archgym}
\bibfield{author}{\bibinfo{person}{Srivatsan Krishnan}, \bibinfo{person}{Amir
  Yazdanbakhsh}, \bibinfo{person}{Shvetank Prakash}, \bibinfo{person}{Jason
  Jabbour}, \bibinfo{person}{Ikechukwu Uchendu}, \bibinfo{person}{Susobhan
  Ghosh}, \bibinfo{person}{Behzad Boroujerdian}, \bibinfo{person}{Daniel
  Richins}, \bibinfo{person}{Devashree Tripathy}, \bibinfo{person}{Aleksandra
  Faust}, {and} \bibinfo{person}{Vijay Janapa~Reddi}.}
  \bibinfo{year}{2023}\natexlab{}.
\newblock \showarticletitle{{ArchGym}: An Open-Source Gymnasium for Machine
  Learning Assisted Architecture Design}. In
  \bibinfo{booktitle}{\emph{Proceedings of the 50th Annual International
  Symposium on Computer Architecture}} \emph{(\bibinfo{series}{ISCA '23})}.
  \bibinfo{publisher}{Association for Computing Machinery}.
\newblock
\href{https://doi.org/10.1145/3579371.3589049}{doi:\nolinkurl{10.1145/3579371.3589049}}


\bibitem[Li et~al\mbox{.}(2025b)]%
        {lemoe}
\bibfield{author}{\bibinfo{person}{Jingyuan Li}, \bibinfo{person}{Jianrong
  Zhang}, \bibinfo{person}{Ye Li}, \bibinfo{person}{Wenbo Yin}, {and}
  \bibinfo{person}{Lingli Wang}.} \bibinfo{year}{2025}\natexlab{b}.
\newblock \showarticletitle{{LEMOE}: {LLM}-Enhanced Multi-Objective Bayesian
  Optimization for Microarchitecture Exploration}. In
  \bibinfo{booktitle}{\emph{Proceedings of the 62nd ACM/IEEE Design Automation
  Conference}} \emph{(\bibinfo{series}{DAC '25})}. \bibinfo{pages}{1--7}.
\newblock
\href{https://doi.org/10.1109/DAC63849.2025.11132704}{doi:\nolinkurl{10.1109/DAC63849.2025.11132704}}


\bibitem[Li et~al\mbox{.}(2025a)]%
        {idse}
\bibfield{author}{\bibinfo{person}{Runkai Li}, \bibinfo{person}{Jia Xiong},
  {and} \bibinfo{person}{Xi Wang}.} \bibinfo{year}{2025}\natexlab{a}.
\newblock \bibinfo{title}{{iDSE}: Navigating Design Space Exploration in
  High-Level Synthesis Using {LLMs}}.
\newblock
\showeprint[arxiv]{2505.22086}~[cs.AR]


\bibitem[Liu et~al\mbox{.}(2024)]%
        {llambo}
\bibfield{author}{\bibinfo{person}{Tennison Liu}, \bibinfo{person}{Nicol\'{a}s
  Astorga}, \bibinfo{person}{Nabeel Seedat}, {and} \bibinfo{person}{Mihaela
  van~der Schaar}.} \bibinfo{year}{2024}\natexlab{}.
\newblock \showarticletitle{Large Language Models to Enhance {Bayesian}
  Optimization}. In \bibinfo{booktitle}{\emph{The Twelfth International
  Conference on Learning Representations (ICLR)}}.
\newblock
\urldef\tempurl%
\url{https://openreview.net/forum?id=OOxotBmGol}
\showURL{%
\tempurl}


\bibitem[Pandit et~al\mbox{.}(2025)]%
        {llmbox}
\bibfield{author}{\bibinfo{person}{Sujay Pandit}, \bibinfo{person}{Akanksha
  Jain}, \bibinfo{person}{Rami Cohen}, \bibinfo{person}{Zhijie Deng},
  \bibinfo{person}{Sagar Karandikar}, \bibinfo{person}{Sagi Perel},
  \bibinfo{person}{Anand Raghunathan}, {and} \bibinfo{person}{Parthasarathy
  Ranganathan}.} \bibinfo{year}{2025}\natexlab{}.
\newblock \showarticletitle{{LLM-Box}: An Agentic Framework for Guided
  Black-Box Optimization in Mapping {LLMs} onto Specialized Hardware
  Accelerators}. In \bibinfo{booktitle}{\emph{Machine Learning for Systems
  (MLForSys) Workshop at NeurIPS}}.
\newblock


\bibitem[Radford et~al\mbox{.}(2021)]%
        {clip}
\bibfield{author}{\bibinfo{person}{Alec Radford}, \bibinfo{person}{Jong~Wook
  Kim}, \bibinfo{person}{Chris Hallacy}, \bibinfo{person}{Aditya Ramesh},
  \bibinfo{person}{Gabriel Goh}, \bibinfo{person}{Sandhini Agarwal},
  \bibinfo{person}{Girish Sastry}, \bibinfo{person}{Amanda Askell},
  \bibinfo{person}{Pamela Mishkin}, \bibinfo{person}{Jack Clark},
  {et~al\mbox{.}}} \bibinfo{year}{2021}\natexlab{}.
\newblock \showarticletitle{Learning Transferable Visual Models From Natural
  Language Supervision}. In \bibinfo{booktitle}{\emph{International Conference
  on Machine Learning (ICML)}}. \bibinfo{pages}{8748--8763}.
\newblock


\bibitem[Radford et~al\mbox{.}(2023)]%
        {whisper}
\bibfield{author}{\bibinfo{person}{Alec Radford}, \bibinfo{person}{Jong~Wook
  Kim}, \bibinfo{person}{Tao Xu}, \bibinfo{person}{Greg Brockman},
  \bibinfo{person}{Christine McLeavey}, {and} \bibinfo{person}{Ilya
  Sutskever}.} \bibinfo{year}{2023}\natexlab{}.
\newblock \showarticletitle{Robust Speech Recognition via Large-Scale Weak
  Supervision}. In \bibinfo{booktitle}{\emph{International Conference on
  Machine Learning (ICML)}}. \bibinfo{pages}{28492--28518}.
\newblock


\bibitem[Sakhuja et~al\mbox{.}(2023)]%
        {spotlight}
\bibfield{author}{\bibinfo{person}{Chirag Sakhuja}, \bibinfo{person}{Zhan Shi},
  {and} \bibinfo{person}{Calvin Lin}.} \bibinfo{year}{2023}\natexlab{}.
\newblock \showarticletitle{Leveraging Domain Information for the Efficient
  Automated Design of Deep Learning Accelerators}. In
  \bibinfo{booktitle}{\emph{IEEE International Symposium on High-Performance
  Computer Architecture (HPCA)}}. \bibinfo{pages}{287--301}.
\newblock
\href{https://doi.org/10.1109/HPCA56546.2023.10071095}{doi:\nolinkurl{10.1109/HPCA56546.2023.10071095}}


\bibitem[Sun et~al\mbox{.}(2019)]%
        {hrnet}
\bibfield{author}{\bibinfo{person}{Ke Sun}, \bibinfo{person}{Bin Xiao},
  \bibinfo{person}{Dong Liu}, {and} \bibinfo{person}{Jingdong Wang}.}
  \bibinfo{year}{2019}\natexlab{}.
\newblock \showarticletitle{Deep High-Resolution Representation Learning for
  Human Pose Estimation}. In \bibinfo{booktitle}{\emph{IEEE/CVF Conference on
  Computer Vision and Pattern Recognition (CVPR)}}.
  \bibinfo{pages}{5693--5703}.
\newblock


\bibitem[Topcuoglu et~al\mbox{.}(2002)]%
        {heft}
\bibfield{author}{\bibinfo{person}{Haluk Topcuoglu}, \bibinfo{person}{Salim
  Hariri}, {and} \bibinfo{person}{Min-You Wu}.}
  \bibinfo{year}{2002}\natexlab{}.
\newblock \showarticletitle{Performance-Effective and Low-Complexity Task
  Scheduling for Heterogeneous Computing}.
\newblock \bibinfo{journal}{\emph{IEEE Transactions on Parallel and Distributed
  Systems}} \bibinfo{volume}{13}, \bibinfo{number}{3} (\bibinfo{year}{2002}),
  \bibinfo{pages}{260--274}.
\newblock
\href{https://doi.org/10.1109/71.993206}{doi:\nolinkurl{10.1109/71.993206}}


\bibitem[Wang et~al\mbox{.}(2025)]%
        {llmdse}
\bibfield{author}{\bibinfo{person}{Hanyu Wang}, \bibinfo{person}{Xinrui Wu},
  \bibinfo{person}{Zijian Ding}, \bibinfo{person}{Su Zheng},
  \bibinfo{person}{Chengyue Wang}, \bibinfo{person}{Neha Prakriya},
  \bibinfo{person}{Tony Nowatzki}, \bibinfo{person}{Yizhou Sun}, {and}
  \bibinfo{person}{Jason Cong}.} \bibinfo{year}{2025}\natexlab{}.
\newblock \bibinfo{title}{{LLM-DSE}: Searching Accelerator Parameters with
  {LLM} Agents}.
\newblock
\showeprint[arxiv]{2505.12188}~[cs.AR]


\bibitem[Yang et~al\mbox{.}(2024b)]%
        {qwen25}
\bibfield{author}{\bibinfo{person}{An Yang}, \bibinfo{person}{Baosong Yang},
  \bibinfo{person}{Beichen Zhang}, \bibinfo{person}{Binyuan Hui},
  \bibinfo{person}{Bo Zheng}, {et~al\mbox{.}}}
  \bibinfo{year}{2024}\natexlab{b}.
\newblock \bibinfo{title}{Qwen2.5 Technical Report}.
\newblock
\showeprint[arxiv]{2412.15115}~[cs.CL]


\bibitem[Yang et~al\mbox{.}(2024a)]%
        {dav2}
\bibfield{author}{\bibinfo{person}{Lihe Yang}, \bibinfo{person}{Bingyi Kang},
  \bibinfo{person}{Zilong Huang}, \bibinfo{person}{Zhen Zhao},
  \bibinfo{person}{Xiaogang Xu}, \bibinfo{person}{Jiashi Feng}, {and}
  \bibinfo{person}{Hengshuang Zhao}.} \bibinfo{year}{2024}\natexlab{a}.
\newblock \showarticletitle{Depth Anything V2}. In
  \bibinfo{booktitle}{\emph{Advances in Neural Information Processing
  Systems}}.
\newblock


\bibitem[Zitzler and Thiele(1999)]%
        {hypervolume}
\bibfield{author}{\bibinfo{person}{Eckart Zitzler} {and}
  \bibinfo{person}{Lothar Thiele}.} \bibinfo{year}{1999}\natexlab{}.
\newblock \showarticletitle{Multiobjective Evolutionary Algorithms: A
  Comparative Case Study and the Strength {Pareto} Approach}.
\newblock \bibinfo{journal}{\emph{IEEE Transactions on Evolutionary
  Computation}} \bibinfo{volume}{3}, \bibinfo{number}{4}
  (\bibinfo{year}{1999}), \bibinfo{pages}{257--271}.
\newblock


\end{thebibliography}

%% ─── Appendix (does NOT count toward 6-page limit) ──────────
\appendix
\section{Appendix: Case Study --- Proposer--Critic Reasoning on HRNet}
\label{app:case_study}

\begin{figure}[!t]
\centering
\begin{minipage}{\columnwidth}\centering
{\footnotesize\textbf{(a)}}\\[-1pt]
\includegraphics[width=\columnwidth]{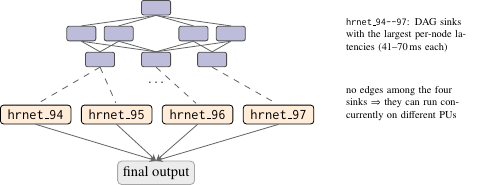}\\[5pt]
{\footnotesize\textbf{(b)}}\\[-1pt]
\includegraphics[width=\columnwidth]{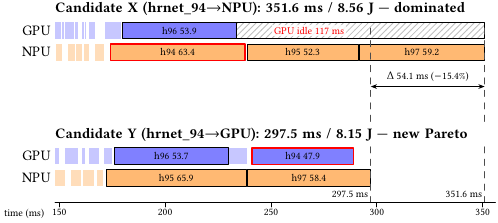}\\[5pt]
{\footnotesize\textbf{(c)}}\\[-1pt]
\includegraphics[width=\columnwidth]{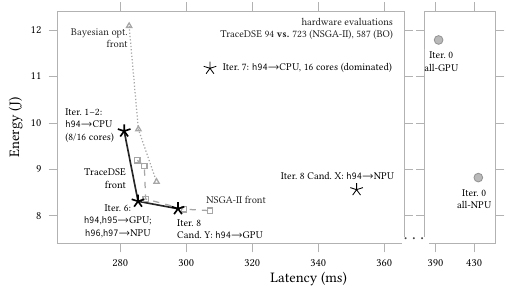}
\end{minipage}\\[-2pt]
\caption{Case study on HRNet. \textbf{(a)}~Terminal fragment of the workload DAG: the four sink nodes are mutually independent and can run concurrently on different PUs. \textbf{(b)}~Execution traces of the two Iteration-8 candidates differing in a single node: Candidate~X (\texttt{hrnet\_94}$\to$NPU) serializes three terminal nodes and idles the GPU for 117\,ms (dominated); Candidate~Y (\texttt{hrnet\_94}$\to$GPU) runs both accelerators concurrently, 54.1\,ms (15.4\%) faster (new Pareto point). \textbf{(c)}~Search trajectory from the Iteration-0 seeds---all-GPU and all-NPU, i.e.\ every node mapped to a single PU---to TraceDSE's final Pareto front (axis break marked by dots). Each labeled point gives the iteration and the mapping decision that defines it; dashed and dotted lines are the NSGA-II and Bayesian-optimization final fronts after their full 12-hour budgets. Throughout, \texttt{h94} abbreviates \texttt{hrnet\_94}.}
\Description{A DAG fragment with four independent terminal nodes, two Gantt charts comparing execution traces of two candidate mappings, and a latency-energy plot showing the search trajectory and Pareto fronts.}
\label{fig:case_study}
\end{figure}

We trace the proposer--critic loop across 12~iterations on HRNet to show how execution-trace feedback drives iterative improvement. HRNet's 98-node DAG ends in the four terminal nodes \texttt{hrnet\_94}--\texttt{hrnet\_97} (Figure~\ref{fig:case_study}(a)): mutually independent sinks that carry the largest per-node latencies in the workload (41--70\,ms each). Their placement therefore largely determines end-to-end latency. Italicized text is quoted verbatim from the run.

\paragraph{Iterations 0--1.}
Before the loop, one-time profiling measured each node's latency and energy on every PU at maximum resources; these priors warm-start the proposer and give the critic a contention-free reference for interpreting traces. In Iteration~1 the proposer first distributed the 98~nodes across all three PUs and reached 283\,ms---28\% better than the all-GPU seed---placing \texttt{hrnet\_94} on the CPU, which the priors rank fastest for it in isolation (40.2\,ms). The trace told a different story: \texttt{hrnet\_94} took 64.2\,ms because the configuration enabled only half the CPU cores (``\textit{\ldots\ CPU has only 8 of 16 cores active, causing \texttt{hrnet\_94} to run slower than expected}''), so the critic recommended the GPU (a predicted 1.54$\times$ speedup). The proposer began testing that placement immediately---once in Iteration~2 and three times in Iteration~3---but spent Iterations~2--5 mainly on the 94~non-terminal nodes, and the terminal assignment did not settle; the front grew to span 281--352\,ms / 8.2--9.8\,J. The GPU placement only became decisive in Iteration~6, once it was paired with the balanced split.

\paragraph{Iteration 6.}
The proposer then assigned two terminal nodes to the GPU and two to the NPU, reasoning that spreading independent work across PUs avoids serializing it; the candidate reached 287\,ms / 8.3\,J, improving both objectives (a new Pareto point). From the trace the critic identified the next bottleneck: \texttt{hrnet\_96} at 57\,ms on the NPU versus a predicted 46.6\,ms on the GPU (1.22$\times$).

\paragraph{Iterations 7--8.}
In Iteration~7 the proposer moved \texttt{hrnet\_94} back to the CPU with all 16 cores active, probing whether CPU placement could reduce GPU energy; every candidate was dominated (307--353\,ms). The critic's trace analysis explained why: contention from other CPU-mapped nodes inflated \texttt{hrnet\_94} to 63.1\,ms, far above its 40.2\,ms contention-free prior (``\textit{Terminal nodes are extremely sensitive \ldots\ CPU is too slow even with all cores active}''), so it recommended moving the node off the CPU and, since the CPU would no longer run terminal nodes, halving its allocation to eight cores.

Of the six Iteration-8 candidates---all adopting the configuration change---four followed the critic and placed \texttt{hrnet\_94} on the NPU, while two placed it on the GPU, the proposer reasoning from the DAG that the four terminal nodes can run concurrently if spread across PUs (``\textit{Extreme Latency Push: Perfectly splits the 4 terminal heads across the accelerators \ldots\ for balanced tail execution}''). Call the two single-node-difference candidates X (\texttt{hrnet\_94}$\to$NPU) and Y (\texttt{hrnet\_94}$\to$GPU); both use identical PU configurations. As Figure~\ref{fig:case_study}(b) shows, X serializes three terminal nodes on the NPU (351.6\,ms, dominated) and loses 54\,ms to Y's concurrent split (297.5\,ms, new Pareto)---far more than the 15.5\,ms per-node difference between the two placements (47.9 vs.\ 63.4\,ms).

\paragraph{Iterations 9--12.}
The critic generalized: ``\textit{An exactly balanced 2/2 split between GPU and NPU is mandatory for sub-300\,ms latency}''---two terminal nodes on each accelerator---and predicted only 5--10\,ms variance across permutations. Iterations~9--10 confirmed this (291--301\,ms); energy-focused configurations in Iterations~11--12 added no Pareto points, and the search terminated since five iterations resulted in hypervolume improvement below the 0.5\% threshold.

\paragraph{Summary.}
The agents contribute complementary evidence: the critic's trace analysis corrects priors that contention makes misleading (40.2\,ms in isolation versus 63--64\,ms in situ) and turns each evaluation into a targeted recommendation, while the proposer adds DAG-level reasoning about concurrency that traces alone do not provide. Together they drive the search to a balanced GPU--NPU terminal split with a reduced CPU allocation, producing the 15.4\% latency reduction cited in Section~\ref{subsec:ablation}.

\end{document}